\documentclass{ifacconf}

\usepackage{graphicx}      
\usepackage{natbib}        

\usepackage[nolist,nohyperlinks]{acronym}
\usepackage{amsmath,amssymb,amsfonts}
\usepackage{booktabs}
\usepackage{makecell}
\usepackage{csquotes}
\usepackage{ifthen}
\usepackage{siunitx} 

\def\colortheme{1}

\newcommand{\imag}{\ensuremath{\mathrm{\imath}}} 
\newcommand{\expAngle}[1]{\mathrm{e}^{\imag #1}}
\newtheorem{remark}{Remark}
\newtheorem{assumption}{Assumption}
 
\usepackage{tikz}
\usetikzlibrary{arrows}
\usetikzlibrary{decorations.pathreplacing,decorations.markings} 
\usetikzlibrary{positioning}
\usetikzlibrary{calc}
\usetikzlibrary{fit}
\usetikzlibrary{backgrounds}

\usepackage[americanvoltages, americaninductors]{circuitikz}
\ctikzset{bipoles/length=0.85cm}
\ctikzset{bipoles/thickness=1}
\usepackage{pgfplots}
\pgfplotsset{compat=1.9}
\usepgfplotslibrary{external}
\usepgfplotslibrary{statistics}
\pgfplotsset{every axis/.append style={semithick,tick style={major tick
            length=4pt,semithick,black}}}

\tikzset{external/system call={pdflatex \tikzexternalcheckshellescape -halt-on-error
        -interaction=batchmode -jobname "\image" "\texsource"}}
\definecolor{hks}{RGB}{199,16,92}
\definecolor{Dunkelgr}{RGB}{21, 56, 36}
\definecolor{Hellblau}{RGB}{80, 149, 200}
\definecolor{Gelbgrue}{RGB}{196, 210, 15}
\definecolor{Hellgrue}{RGB}{74, 172, 150}
\definecolor{Goldgelb}{RGB}{234, 195, 114}

\definecolor{vgRed}{RGB}{193, 48, 24}
\definecolor{vgOrange}{RGB}{243, 111, 19}
\definecolor{vgYellow}{RGB}{235, 203, 56}
\definecolor{vgGreen}{RGB}{162, 185, 105}
\definecolor{vgLightBlue}{RGB}{13, 149, 188}
\definecolor{vgDarkBlue}{RGB}{6, 56, 81}

\definecolor{cbBlue}{HTML}{0072B2}
\definecolor{cbOrange}{HTML}{E69F00}
\definecolor{cbGreen}{HTML}{009E73}

\pgfplotsset{every axis/.append style={semithick,tick style={major tick
            length=4pt,semithick,black}}}

\pgfkeys{/pgfplots/x axis shift down/.style={
        x axis line style={yshift=-#1},
        xtick style={yshift=-#1},
        tick align=outside,
        xticklabel shift={#1}}}

\pgfkeys{/pgfplots/y axis shift left/.style={
        y axis line style={xshift=-#1},
        ytick style={xshift=-#1},
        yticklabel shift={#1},
        scaled y ticks = false,
        tick align=outside,
        y tick label style={/pgf/number format/fixed,
        }
    }
}

\pgfplotsset{myPlot/.style={%
        width=8cm,
        height=3.5cm,
        line width = 0.7pt,
        separate axis lines,
        axis x line*=bottom,
        x axis shift down = 3pt,
        enlarge x limits=false,
        axis y line*=left,
        y axis shift left = 6pt,
        enlarge y limits={abs=.25pt},
        enlarge x limits={abs=.25pt},
    }
}

\begin{document}
\begin{acronym}
    \acro{ml}[ML]{machine learning}
    \acro{nn}[NN]{neural network}
    \acro{mlp}[MLP]{multilayer perceptron}
    \acro{lstm}[LSTM]{long short-term memory}
    \acro{tcn}[TCN]{temporal convolutional network}
    \acro{ho}[HO]{hyperparameter optimization}
    \acro{ivp}[IVP]{initial value problem}
    \acro{ode}[ODE]{ordinary differential equation}
    \acro{node}[NODE]{neural ordinary differential equation}
    \acro{dnode}[D-NODE]{data-controlled neural ordinary differential equation}
    \acro{anode}[A-NODE]{augmented neural ordinary differential equation}
    \acro{pinn}[PINN]{physics informed neural network}
    \acro{nhnn}[NHNN]{nearly hamiltonian neural network}
    \acro{sindy}[SINDy]{sparse identification of nonlinear dynamics}
    \acro{dp5}[DOPRI5]{fifth-order Dormand-Prince}
    \acro{rk4}[RK4]{fourth-order Runge-Kutta}
    \acro{der}[DER]{distributed energy resource}
    \acro{sg}[SG]{synchronous generator}
    \acro{gfi}[GFI]{grid-forming inverter}
    \acro{smib}[SMIB]{single-machine-infinite-bus}
    \acro{nhnn}[NHNN]{nearly hamiltonian neural network}
    \acro{rmse}[RMSE]{root-mean-square-error}
    \acro{mse}[MSE]{mean squared error}
    \acro{relu}[ReLU]{rectified linear unit}
    \acro{gelu}[GELU]{Gaussian error linear unit}
    \acro{silu}[SiLU]{sigmoid linear unit}

    \acro{io}[I/O]{input/output}

    \acro{ieee9}[IEEE9]{IEEE 9 bus system}
    \acro{ieee30}[IEEE30]{IEEE 30 bus system}
    \acro{ieee39}[IEEE39]{IEEE 39 bus system}
\end{acronym}

\begin{frontmatter}
 
\title{Augmented Neural Ordinary Differential Equations for Power System Identification} 

  
\author[First]{Hannes M. H. Wolf} 
\author[First]{Christian A. Hans} 
  
\address[First]{Automation and Sensorics in Networked Systems, University of Kassel, Germany (e-mail: \{h.wolf, hans\}@uni-kassel.de)}

\begin{abstract}                
Due the complexity of modern power systems, modeling based on first-order principles becomes increasingly difficult. 
As an alternative, dynamical models for simulation and control design can be obtained by black-box identification techniques. 
One such technique for the identification of continuous-time systems are \aclp{node}. 
For training and inference, they require initial values of system states, such as phase angles and frequencies. 
While frequencies can typically be measured, phase angle measurements are usually not available. 
To tackle this problem, we propose a novel structure based on \aclp{anode}, learning latent phase angle representations on historic observations with \aclp{tcn}. 
Our approach combines state-of-the art deep learning techniques, avoiding the necessity of phase angle information for the power system identification. 
Results show, that our approach clearly outperforms simpler augmentation techniques.
\end{abstract}


\end{frontmatter}

\section{Introduction}
In power systems engineering, dynamical models are crucial for simulation, analysis and control design. 
Traditionally, such models are obtained using first order principles. 
In future power systems with a large number of distributed renewable and storage units, this approach is becoming increasingly difficult. 
At the same time, ongoing digitalization and widerspread deployment of sensors in powergrids offer a vast amount of measurements that open the door for data-driven system identification.
Here, dynamical models are obtained even without knowledge of the underlying processes, which can be particularily useful for large scale future power systems. 

A modern technique for black-box system identification are neural \acp{ode}. 
They were first introduced by \cite{chen_neural_2018} and allow to represent the right hand side of an \ac{ode} as a \ac{nn}. 
Since then, several extensions to neural \acp{ode} (\acsp{node}) have been proposed. \acused{node}
\cite{massaroli_dissecting_2020} introduced data-controlled \acp{node} which allow to incorporate external control and disturbance inputs. 
Furthermore, \cite{dupont_augmented_2019} introduced augmented \acp{node} that enable latent representations of unmeasured states. 
In the power system domain, several studies have employed \acp{node} to identify components or entire power systems. 
\cite{aryal_neuralps_2023} employed \acp{node} to identify linearized \ac{io} frequency dynamics of droop-controlled \acp{sg}.  
However, coupling between \acp{sg} was not considered. 
\cite{zhang_learning_2024} used \acp{node} to identify the frequency dynamics of the entire IEEE 118 bus system. 
They focused on generator angle and frequency trajectories under fault scenarios. 
\cite{wolf_ident_2025} considered coupled dynamics of droop-controlled \acp{gfi}. 
Using desired frequency and voltage setpoints as control input, voltage and frequency dynamics were identified with \acp{node} and compared to symbolic regression. 
All of the aforementioned studies assumed that phase angles and frequency of all units are available. 
Opposed to this, \cite{xiao_feasibility_2023} used \acp{node} to identify the dynamics of power system components from portal measurements. 
They emphasize scenarios where the system state could not be measured and employ autoencoders to obtain initial states from available values of a single time instant. 

Typically, observations of a single time instant do not suffice to uniquely determine a system state. 
This is why \cite{masti_autoenc_2018}, as well as \cite{forgione_learning_initial_2022_arx} use a time series of historic observations of system inputs and outputs to learn the initial state for neural state space models using deep autoencoders and \acp{lstm}, respectively. 
In this paper we build on this idea and extend it for augmented \acp{node} in order to learn latent representations for non-measured system states, i.e., unknown phase angles. 
Our contributions are as follows: 
1) We combine state-of-the-art deep learning frameworks for system identification. 
In detail, we propose a novel structure that, for the first time, employs \acp{tcn} to augment unmeasured states in \acp{node}. 
2) We learn voltage and frequency dynamics of entire power systems with our novel structure, without using phase angle information. 
This is crucial, as phase angle measurements can only be obtained with costly measurement equipment, such as phasor measurement units.
3) We conduct large scale hyperparameter optimizations to learn models for three power systems of different size. 
In a numerical case study, we show that our approach outperforms augmented \acp{node} with simpler augmentation techniques. 

The paper is structured as follows. 
In Section~\ref{sec:preliminaries}, we define our notation and introduce fundamental \ac{nn} structures. 
Section~\ref{sec:modeling} discusses power system models. 
In Section~\ref{sec:nodes_for_system_identification}, we describe the fundamentals of \acp{node}, bridge the gap from \ac{ml} to system identification and introduce our novel approach. 
The case study and the results are presented in Section~\ref{sec:case_study}. 
Section~\ref{sec:conclusion} concludes the paper.

\section{Preliminaries}
\label{sec:preliminaries}

\subsection{Notation}
\label{sec:notation}

We denote the set of real numbers by $\mathbb{R}$, the set of non-negative real numbers by $\mathbb{R}_{\geq 0}$, the set of strictly positive real numbers by $\mathbb{R}_{> 0}$, the set of nonegative integers $\mathbb{N}$ and the set of complex numbers by $\mathbb{C}$. 
An element $z \in \mathbb{C}$ is $z = a + \imag b$ where $\imag$ is the imaginary unit, $a \in \mathbb{R}$ is the real part and $b \in \mathbb{R}$ is the imaginary part. 
We donete datasets as $\mathcal{D} = \{(\xi_n, \eta_n)\}_{n=1}^N$ where the $n$-th sample $(\xi_n, \eta_n)$ consists of features $\xi_n \in \mathbb{R}^{n_\xi}$, i.e., the inputs to a \ac{ml} model, and targets $\eta_n \in \mathbb{R}^{n_\eta}$, i.e., the expected outputs for features $\xi_n$. 
\subsection{\Aclp{nn}}
\label{sec:neural_networks}

\subsubsection{\Aclp{mlp}}
\label{sec:mlp}
A \ac{mlp} is a feedforward \ac{nn} with $n_\xi \in \mathbb{N}$ inputs and $n_\eta \in \mathbb{N}$ outputs consisting of $L \in \mathbb{N}$ fully connected layers of neurons. 
Each layer $l \in \{1, \dots , L\}$ contains $N_{f,l}$ neurons and applies an affine transformation, which is, except for the last layer, followed by a nonlinear activation function $\sigma$. 
The input $\xi \in \mathbb{R}^{n_\xi}$ is mapped to the output with 
\begin{equation}
    \label{eq:mlp}
    f_\theta(\xi) = 
     W_{L} \sigma(W_{L-1} \dots \sigma(W_1 \,  \xi + b_1) \dots + b_{L-1})+ b_{L},
\end{equation}
where $W_i \in \mathbb{R}^{N_{f,i} \times N_{f,i-1}}$ are weights and $b_i \in \mathbb{R}^{N_{f,i}}$ are biases, i.e., the parameters $\theta$ of the \ac{nn}, with $N_{f,0} = n_\xi$ and $N_{f,L} = n_\eta$. 

\subsubsection{\Aclp{tcn}} 
\label{sec:tcn}
\Acp{tcn} \citep{bai_tcn_2018_arx} are based on causal convolutions, which allow to capture temporal dependencies in sequences. 
Consider the sequence $\xi = \{\xi(t_1), \allowbreak \xi(t_2), \allowbreak \dots, \xi(t_T)\}$ of length $T$ with $\xi(t_k) \in \mathbb{R}^I$, where $I \in \mathbb{N}$ is the number of channels.  
When using the entire sequence as an input to a causal 1D-convolutional layer, we obtain a sequence $\eta$ at the output of same length but with $O \in \mathbb{N}$ channels, i.e., $\eta(t_k) \in R^O$.
At channel $o \in [1, O]\subset \mathbb{N}$ the output at time $t_k$ is
\begin{equation}
    \label{eq:tcn_conv}
    \eta_o(t_k) = b^{o} + \sum_{i=1}^{I} \sum_{j=0}^{K-1}  W_{i,j}^{o} \, \xi_i(t_{k - d  j})
\end{equation}
with kernel $W^{o} \in \mathbb{R}^{K \times I}$ and bias $b^{o} \in \mathbb{R}$. 
The dilation $d \in \mathbb{N}$ is the temporal spacing between elements of the inputs in the convolution. 

We use residual blocks $h_b, \, b \in [0, B-1] \subset \mathbb{N}$ as described in \cite{bai_tcn_2018_arx}, where one residual block consists of two convolutional layers \eqref{eq:tcn_conv}, each with $O_b$ hidden channels followed by a \ac{relu}. 
The output sequence of such a residual connection for input $\xi$ is
\begin{equation}
    \eta = \xi + h_b(\xi).
\end{equation}
Finally, \acp{tcn} consist of $B$ consecutive residual connections with an exponentially growing dilation $d_b = 2^b$. 
Thus, there are $\sum_{l=0}^{B-1} 2\, O_b$ kernels and biases, which are the parameters of the \ac{tcn}.
The effective history of the \ac{tcn} is
\begin{equation}
    \label{eq:tcn_recep}
    R = 1 + 2 (K - 1) (2^B - 1),
\end{equation}
i.e., in order to utilize all $\xi(t_k)$ of input $\xi$ for $\eta(t_T)$ of output $\eta$, $R \geq T$ must hold.

\section{Modeling}
\label{sec:modeling}

In what follows, we derive the mathematical model of the power system components. 
We assume balanced three-phase grids and represent electrical quantities using complex numbers in a rotating coordinate system. 
The network, as well as the dynamic models of the units mainly follow \cite{schiffer_stability_2015}. 

\subsection{Network model}
\label{sec:network_model}
The electrical network is modelled as a graph $\mathcal{G} = (\mathcal{N}, \mathcal{E},y)$, where $\mathcal{N}$ is the set of $|\mathcal{N}|$ nodes, $\mathcal{E} \subseteq \mathcal{N} \times \mathcal{N}$ the set of edges and $y: \mathcal{E} \rightarrow \mathbb{C}, \quad y((i,j)) = {y}_{ij} = g_{ij} + \imag b_{ij}$ is a weighting function returning the admittance of the edge $(i,j)$. 
Here $g_{ij} \in \mathbb{R}_{\geq 0}$ and $b_{ij} \in \mathbb{R}$ denote the conductance and the susceptance, respectively. 
In this context, self edges $(i,i) \in \mathcal{E}$ are associated with parallel shunt admittances to the ground and $(i,j) \in \mathcal{E}, \, i\neq j$ with series admittances between distinct nodes.
The set of nodes adjacent to $i \in \mathcal{N}$ is referred to as $\mathcal{N}_i \subset \mathcal{N}$.
Each node $i \in \mathcal{N}$ is associated with a voltage phasor $\tilde{v}_i = v_i \expAngle{\varphi_i} \in \mathbb{C}$, where $v_i \in \mathbb{R}_{\geq 0}$ is the voltage amplitude and $\varphi_i \in \mathbb{R}$ the voltage phase angle. 
Furthermore, each node is equipped with a local $dq$-coordinate system rotating with $\omega_i \in \mathbb{R}$, whose $q$-axis is aligned with the voltage phasor. 
We express phase angle differences of these coordinate systems to a global reference $dq$-system as $\delta_{i} = \varphi_i - \varphi_{\text{ref}}$, and thus $\dot{\delta}_i = \omega_i - \omega_{\text{ref}}$.

Consider the admittance matrix $Y \in \mathbb{C}^{|\mathcal{N}| \times |\mathcal{N}|}$ \citep{kundur_power_1994} constructed from $y_{ij}, \forall (i,j) \in \mathcal{E}$, as well as the vectors $\tilde{v} \in \mathbb{C}^{|\mathcal{N}|}$ and ${c} \in \mathbb{C}^{|\mathcal{N}|}$ collecting the node voltages and injected currents of each node, respectively.  
We can compute the injected complex power for all nodes with ${s} = \tilde{v} {c}^* = \tilde{v} (Y\, \tilde{v})^*$. 
Using ${s}_i= p_i + \imag q_i$, where $p_i \in \mathbb{R}$ is the active and $q_i \in \mathbb{R}$ the reactive power, as well as $\delta_{ij} = \delta_i - \delta_j$, we obtain 
\begin{align}
\label{eq:power_injections}
    p_{i} &=  \, v_i^2 \, g_{ii} + v_i\textstyle\sum\limits_{j \in \mathcal{N}_i} v_i g_{ij} - v_j\left(g_{ij} \cos (\delta_{ij}) + b_{ij} \sin (\delta_{ij})\right), \notag\\
    q_{i} &= - v_i^2 \, b_{ii} - v_i \textstyle\sum\limits_{j \in \mathcal{N}_i} v_i b_{ij} + v_j\left(g_{ij} \sin (\delta_{ij}) - b_{ij} \cos (\delta_{ij})\right).
\end{align}

\subsection{Unit models}
\label{sec:generator_model}
We consider setups, where each node $i \in \mathcal{N}$ is equipped with a droop-controlled \ac{gfi} or a \ac{sg}. 
The set of nodes with \acp{gfi} is denoted by $\mathcal{N}_{\text{gfi}}$ and the set of nodes with \acp{sg} by $\mathcal{N}_{\text{sg}}$. 
We proceed to derive state models for both of them. 

\begin{remark}
    In future power systems, \acp{gfi} are taking over parts of the ancilliary services that are provided by SGs in traditional power systems. 
    They can, e.g., operate together with a storage unit providing positive or negative active power to the grid.
\end{remark}
\begin{remark}
    It is possible to consider power systems containing a set of passive nodes $\mathcal{N}_{\text{pas}}$ that do not have a \ac{gfi} or an \ac{sg} connected, i.e., $\mathcal{N} = \mathcal{N}_{\text{gfi}} \cup \mathcal{N}_{\text{sg}} \cup \mathcal{N}_{\text{pas}}$. 
    Such systems can also be modelled using our methods by transforming them into the presented ones by applying Kron reduction \citep{dorfler_kron_2013}. 
    This maintains the dynamical behavior and thus the resulting state trajectories at nodes $i \in\mathcal{N}_{\text{gfi}} \cup \mathcal{N}_{\text{sg}}$. 
    In practice, system identification would be performed on data obtained from nodes $i \in \mathcal{N}_{\text{gfi}} \cup \mathcal{N}_{\text{sg}}$ without requiring a Kron-reduction step beforehand.
\end{remark}

\subsubsection{\Aclp{gfi}}
\label{sec:grid_forming_inverter}
Droop-controlled \acp{gfi} change frequency and voltage amplitude according to deviations in active and reactive power. 
Following standard practice, we assume these adjustments take place instantaneously \citep{schiffer_conditions_2014}, i.e., 
\begin{subequations}
\label{eq:droop_laws}
\begin{align}
    \label{eq:frequency_droop_law}
    \omega_i&= \omega_i^d - k_i^p (p_i^m - p_i^d), \\
    \label{eq:voltage_droop_law}
    v_i  &= v_i^d - k_i^q (q_i^m - q_i^d).
\end{align}
\end{subequations}
where $p_i^d, \, q_i^d \in \mathbb{R}$ denote the active and reactive power setpoint, respectively. 
Moreover, $\omega_i^d, \,  v_i^d \in \mathbb{R}_{>0}$ are the frequency and voltage amplitude setpoint, respectively, and $k_i^p, \, k_i^q \in \mathbb{R}_{>0}$ are droop gains. 
We assume that active and reactive power measurements, $p_i^m$ and $q_i^m \in \mathbb{R}$, are processed using a first-order low-pass filter with time contant $\tau_i \in \mathbb{R}_{>0}$ \citep{schiffer_conditions_2014}, i.e., 
\begin{subequations}
\label{eq:power_measurement}
\begin{align}
    \label{eq:active_power_measurement}
    \tau_{i} \dot{p}_i^m &= - p_i^m + p_i, \\
    \label{eq:reactive_power_measurement}
    \tau_{i} \dot{q}_i^m &= - q_i^m + q_i.
\end{align} 
\end{subequations}
Combining \eqref{eq:droop_laws} and \eqref{eq:power_measurement} yields the \acp{gfi}' state model at node $i$, i.e.,
\begin{subequations}
\label{eq:gfi_model}
\begin{align}
    \dot{\delta}_{i} &= \omega_{i}^d - k_i^p (p_i^m - p_i^d) - \omega_{ref}, \label{eq:overall_theta} \\
    \dot{p}_i^m &= \frac{1}{\tau_{i}} (- p_i^m + p_{i}), \\
    \dot{v}_i &= \frac{1}{\tau_i} ( -v_i + v_i^d - k_i^q (q_{i} - q_i^d )) .
\end{align}
\end{subequations}

\subsubsection{\Aclp{sg}}
\label{sec:synchronous_generator}
The frequency dynamics of \acp{sg} are modelled using the swing equation \citep{schiffer_stability_2015}, i.e.,
\begin{equation}
    \label{eq:swing_equation}
    m_i \dot{\omega}_i = - k^d_i (\omega_i - \omega_{d}) + p_i^m - p_i,
\end{equation}
where $m_i \in \mathbb{R}_{>0}$ is the inertia constant, $k^d_i \in \mathbb{R}_{>0}$ the damping coefficient and $p_i^m \in \mathbb{R}_{>0}$ is the mechanical power. 
Droop-controlled \acp{sg} change the mechanical power depending on frequency deviations \citep{schiffer_stability_2015}, i.e., 
\begin{equation}
    \label{eq:sg_droop_law}
    p_i^m = p_i^d - \dfrac{1}{k_i^p}(\omega_i - \omega^d).
\end{equation} 
We assume that this change in mechanical power takes place instantaneously with no delay of the govenor. 
Additionally, we consider a similar voltage droop law as for the \acp{gfi}. 
To that end, we assume that the automatic voltage regulator is fast and its dynamics can be neglected such that we can set the voltage at the terminals directly.
Combining \eqref{eq:voltage_droop_law}, \eqref{eq:reactive_power_measurement}, \eqref{eq:swing_equation} and \eqref{eq:sg_droop_law}, we obtain the \acp{sg}' state model at node $i$, i.e., 
\begin{subequations}
\label{eq:sg_model}
\begin{align}
    \dot{\delta}_{i} &= \omega_i - \omega_{ref}, \\
    \dot{\omega}_i &= \frac{1}{m_i} \big(-\frac{1}{k_i^p}(\omega_i - \omega^d) + p_i^d - p_i \big), \\
    \dot{v}_i &= \frac{1}{\tau_i} ( -v_i + v_i^d - k_i^q (q_{i} - q_i^d )) .
\end{align}
\end{subequations}
Note that damping terms have already been included in the droop gain $k_i^p$ \citep{schiffer_stability_2015}.

Combining \eqref{eq:power_injections}, \eqref{eq:gfi_model} and \eqref{eq:sg_model} for all nodes in the grid allows us to derive the overall state model
\begin{subequations}
    \label{eq:nonlinear_state_model}
\begin{align}
    \dot{x}(t) &= f(t,x(t),u(t)) \\
    y(t) &= g(x(t))
\end{align}
\end{subequations}
with state $x(t) \in \mathbb{R}^{n_x}, \, n_x = 3|\mathcal{N}|$ at time $t$, control input $u(t)  \in \mathbb{R}^{n_u}, \, n_u = |\mathcal{N}|$ and output $y(t)  \in \mathbb{R}^{n_y}, \, n_y = 2|\mathcal{N}|$.  
Here, $x$ contains the states, i.e., ${\delta}_{i}(t), p^m_i(t), v_i(t)$ for \acp{gfi} and ${\delta}_{i}(t), \omega_i(t), v_i(t)$ for \acp{sg}, of all units $i \in \mathcal{N}$. 
As control inputs we choose the active power setpoints, i.e., $u(t) = [p_1^d(t) \quad  \cdots \quad  p_N^d(t) ]^T$. 
The states $v_i(t), \omega_i(t), \, \forall i \in \mathcal{N}_{\text{sg}}$ and $v_i(t), p^m_i(t), \, \forall i \in \mathcal{N}_{\text{gfi}}$ are outputs, i.e., $\delta_i(t), \, \forall i \in \mathcal{N}$ is not available to the outside. 


\section{Neural ordinary differential equations}
\label{sec:nodes_for_system_identification}

\begin{figure*}
    \centering
    \input{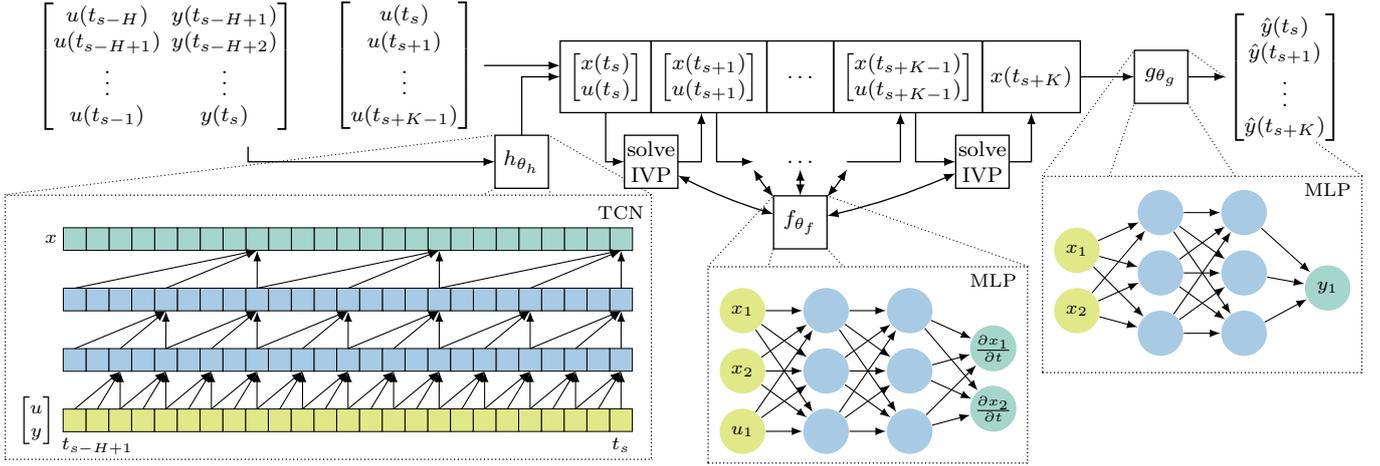}
    \caption{\Ac{ml} model for system identification. Note that all depictions of the \acp{nn} are simplified. For the \ac{tcn}, e.g., we omit the residual blocks and show only convolutional layers instead.}
    \label{fig:neural_ode_stepper}
\end{figure*}

Since a large number of systems can be modeled in the form of \eqref{eq:nonlinear_state_model}, \ac{node}s are a natural choice for black-box system identification. 
Generally speaking, our goal is to approximate the \ac{io} behaviour of \eqref{eq:nonlinear_state_model} with \acp{node}. 

\ac{node}s \citep{chen_neural_2018} represent \acp{ode} where the right hand side is a \ac{nn} $f_{\theta_f}$ with parameters $\theta_{f} \in \mathbb{R}^{n_{\theta_f}}$, i.e.,
\begin{equation}
    \label{eq:neural_ode}
    \dot{x}(t) = f_{\theta_f} (t, x(t)), \quad x(t_{s}) = x_{0}
\end{equation}
Given the initial state $x_{0}$ at time $t_s$, we can use a numerical solver to find a solution to the \ac{ivp} associated with \eqref{eq:neural_ode} at time $t_{e} > t_s$. 
Here, $x_0$ is contained in the features $\xi$, i.e., the inputs to the \ac{ml} model. 

To represent \eqref{eq:nonlinear_state_model}, we use the more general \ac{node} model formulated by \cite{massaroli_dissecting_2020}, i.e., 
\begin{subequations}
\label{eq:data_controlled_neural_ode}
\begin{align}
    \dot{x}(t) &= f_{\theta_f} (t, x(t), e_{\theta_e}(\xi)), \\
    \quad x(t_{s}) &= h_{\theta_h}(\xi), \\
    y(t) &= g(x(t)).
\end{align}
\end{subequations}
Here, the solution to the \ac{ivp} is 
\begin{align}
    \hat{y}(t) &= g_{\theta_g}\big(h_{\theta_h}(\xi) + \int_{t_{s}}^{t_{e}} f_{\theta_f}(t, x(t), e_{\theta_e}(\xi)) dt\big).
\end{align}
Model \eqref{eq:data_controlled_neural_ode} extends \eqref{eq:neural_ode} by three concepts. 
The first one ist called augmentation and was first introduced by \cite{dupont_augmented_2019}. 
In \acp{anode}, $\xi$ is elevated into a higher dimensional space to increase the \acp{node}' expressiveness. 
Possible augmentation techniques are zero-augmentation $x(t_s) = [\xi^T ~ 0^T]^T$ \citep{dupont_augmented_2019}, input-layer augmentation $x(t_s) = h_{\theta_h}(\xi)$ or partial input-layer augmentation $x(t_s) = [\xi^T ~ h^T_{\theta_h}(\xi)]^T$ \citep{massaroli_dissecting_2020}. 
The second concept is called \acp{dnode}, and was first introduced by \cite{massaroli_dissecting_2020}.
Here, the right hand side is conditioned on $\xi$.
This can be done in an explicit or an implicit fashion. 
For the latter, an additional embedding function $\e_{\theta_e}$ is employed. 
Lastly, a function $g_{\theta_g}$ is employed to map the state space to the output space. 

From a control perspective, these extensions come naturally. 
However, the terminology is often different: 
$\xi$ typically contains known control inputs $u$, as well as past and present measurements of $y$, but often no entire state measurements.  
Augmentation means finding $x_0$ from information contained in $\xi$.
Furthermore, \acp{dnode} allow us to incorporate constant control inputs. 
Then, $\e_{\theta_e}(\xi) = u(t_s)$ is just an extraction of $u(t_s)$ from $\xi$ rather than an embedding.  

We aim to approximate \eqref{eq:nonlinear_state_model} with \eqref{eq:data_controlled_neural_ode}. 
In what follows, we discuss different parts of the \ac{ml} model which is shown in Figure~\ref{fig:neural_ode_stepper}. 
First, we describe how to obtain an initial state from a sequence of past observations with a \ac{tcn} $h_{\theta_h}$ in Section~\ref{sec:augmentation}. 
In Section~\ref{sec:system_dynamics_model}, we define the \acp{mlp}  $f_{\theta_f}$ and $g_{\theta_g}$ used to model the dynamics. 
Sections~\ref{sec:time_varying_inputs} and \ref{sec:training_procedure} discuss how to step the model forward in time and train it, respectively. 
We start with some assumptions. 

\begin{assumption}
    We restrict ourselves to time-invariant systems and drop the argument $t$ in $f_{\theta_f}$.
\end{assumption}
\begin{assumption}
    All quantities are sampled at discrete time instants $t_0, t_1, \dots, t_s, \dots, t_{T}$ with constant sampling time $\Delta t = t_{s+1} - t_s$. 
\end{assumption}
\begin{assumption}
    The control input $u(t)$ is piecewise constant over each sampling interval, i.e., $u(t) = u(t_{s})$ for all $t \in [t_{s}, t_{s+1})$. 
\end{assumption}

\subsection{Obtaining an initial state with $h_{\theta_h}$}
\label{sec:augmentation}
For system identification, augmentation is crucial since the dimension of the output space is typically smaller than the dimension of the state space. 
\cite{rahman_neural_2022} propose input-layer augmentation with $x(t_{s}) = h_{\theta_h}(u(t_{s}), y(t_{s}))$.  
In practice $x(t_s)$ can usually not be uniquely determined from $u(t_{s})$ and $y(t_{s})$ at a single time instant.
However, under certain conditions this is possible when $H \geq n_x$ past observations are used \citep{forgione_learning_initial_2022_arx}. 
Therefore, we learn a latent representation $x(t_{s})$ of the actual system state from a history of $H$ observations instead. 
To the best of our knowledge, for the first time \acp{tcn} (see Section~\ref{sec:tcn}) are employed for this task in this paper. 

The methodology is sketeched in the lower left corner of Figure~\ref{fig:neural_ode_stepper}. 
Each $u_i$ and each $y_i$ corresponds to an input channel of the \ac{tcn}. 
With the convolution \eqref{eq:tcn_conv} and \ac{relu} activations they are mapped to hidden channels in the residual blocks. 
Eventually, after the last residual block we obtain a sequence for $x(t_{s-H+1}), \dots, x(t_{s})$ and extract $x(t_{s})$. 
The number of residual blocks $B$ is chosen such that $R \geq H$, where $R$ is obained according to \eqref{eq:tcn_recep}. 
We abuse the notation of the argument of $h_{\theta_h}$ to match the interface of the \ac{tcn} and write 
\begin{equation}
    h_{\theta_h}(\xi) = h_{\theta_h} \left( \begin{bmatrix} u(t_{s-H}) & \ldots & u(t_{s-1}) \\ y(t_{s-H+1}) & \ldots & y(t_{s}) \end{bmatrix} \right).
\end{equation} 
\begin{remark}
    If the output can be partitioned such that $y(t) = [{x^o}^T(t) ~ \tilde{y}^T(t)]^T$ where $x^o$ the part of $x$ that is directly available and $\tilde{y}$ are the remaining outputs, we can also opt for a partial augmentation $x(t_{s}) = [{x^o}^T(t_{s})~ h^T_{\theta_h}(\xi)]^T$. 
\end{remark}

\subsection{Modeling the dynamics with $f_{\theta_f}$ and $g_{\theta_g}$} 
\label{sec:system_dynamics_model}

Let us now discuss $f_{\theta_f}$ and $g_{\theta_g}$ that approximate the \ac{io} dynamics. 
We use \acp{dnode} as described beforehand and condition $f_{\theta_f}$ on the control inputs. 
Thus, we model $f_{\theta_f}$ as a \ac{nn} with inputs $u(t_{s})$ and $x(t)$, as well as outputs $\dot{x}(t)$ using an \ac{mlp} (see Section~\ref{sec:mlp}). 
Similarily, we can use an \ac{mlp} $g_{\theta_g}$ for the output mapping . 
Figure~\ref{fig:neural_ode_stepper} sketches their use in the \ac{ml} model. 
Combining aforementioned concepts and assuming constant control inputs over $[t_s, t_e)$, the model reads
\begin{subequations}
\label{eq:historic_anode}
\begin{align}
    \dot{x}(t) &= f_{\theta_f}(x(t), u(t_{s})), \\
     x(t_{s}) &= h_{\theta_h}  \left( \begin{bmatrix} u(t_{s-H}) & \ldots & u(t_{s-1}) \\ y(t_{s-H+1}) & \ldots & y(t_{s}) \end{bmatrix} \right) \\ 
    \hat{y}(t) &= g_{\theta_g}(x(t)).
\end{align}
\end{subequations}
For the power system problem at hand, we assume that $y$ is a measured part of the system state $x$, i.e., $g_{\theta_g}(x) = x^{o}$. 
Accordingly, we preserve the physical meaning of the measured states in the \ac{node} by choosing a partial augmentation of the initial state and obtain
\begin{subequations}
\label{eq:partial_historic_anode}
\begin{align}
    \dot{x}(t) &= f_{\theta_f} (t, x(t), u(t_{s})), \\
     x(t_{s}) &= \begin{bmatrix}
        x^o(t_{s}) \\
        h_{\theta_h}  \left( \begin{bmatrix} u(t_{s-H}) & \ldots & u(t_{s-1}) \\ x^o(t_{s-H+1}) & \ldots & x^o(t_{s}) \end{bmatrix} \right)
    \end{bmatrix}, \label{eq:partial_historic_augmentation}\\ 
    \hat{y}(t) &= x^o(t).
\end{align}
\end{subequations}

\subsection{Moving the model forward in time}
\label{sec:time_varying_inputs}

In closed-loop simulations, $u(t)$ typically varies over time. 
Thus, an \ac{ivp} with initial time $t_{s}$ might also contain stepwise changes in $u$ over the time horizon $[t_{s}, t_{e})$.

Considering $t_{s}, \allowbreak t_{s+1}, \allowbreak \dots, \allowbreak t_{s+K} \in [t_{s}, t_{e})$, where $K$ is the prediction horizon, we proceed as depicted in Figure~\ref{fig:neural_ode_stepper}.
For given $H$ past measurements of $u$ and $y$, we augment the system state at time $t_{s}$ using $h_{\theta_h}$ and solve the \ac{ivp} up to time $t_{s+1}$ in order to obtain $x(t_{s+1})$. 
For the subsequent time intervals $[t_{s+k}, t_{s+k+1}), \, k = 1, \dots, k = K-1$, we do not need augmentation.
Instead, we can use the solution $x(t_{s+k})$ as the initial state for the \acp{ivp} and condition $f_{\theta_f}$ on $u(t_{s+k})$.
Thereby, we repeatedly solve an \ac{ivp} to step the model forward in time and obtain predictions $\hat{y}(t_{s+1}), \dots, \hat{y}(t_{s+K})$. 

\subsection{Training the model}
\label{sec:training_procedure} 
We train the model on the dataset $\mathcal{D} = \{(\xi_n, \eta_n)\}_{n=1}^N$. 
From the different parts explained in Sections~\ref{sec:augmentation}--\ref{sec:time_varying_inputs}, we end up with features and targets of the form 
\begin{align*}
 \xi_n &= [u^T(t_{s_n-H})  \cdots  u^T(t_{s_n+K}) ~ y^T(t_{s_n-H+1}) \cdots   y^T(t_{s_n})]^T, \\
 \eta_n &= [y^T(t_{s_n+1})\,  \cdots \,y^T(t_{s_n+K})]^T,
\end{align*}
and minimize the mean square error between the targets $\eta_n$ and the predictions resulting from $\xi_n$, i.e.,
\begin{equation}
    L = \dfrac{1}{N K}\sum_{n=1}^{N} \sum_{k=1}^{K} ||y(t_{s_n+k}) - \hat{y}_n(t_{s_n+k})||_2^2.
\end{equation}
To obtain $\partial L / \partial \theta$, we use the "discretize-then-optimize" approach \citep{kidger_neural_2022}, propagating gradients through solver operations.  


\section{Case Study}
\label{sec:case_study}
In what follows, we employ the methods from Section~\ref{sec:nodes_for_system_identification} to identify systems of the form described in Section~\ref{sec:modeling}. 
In particular, we consider three grids of different size: the \ac{ieee9}, the \ac{ieee30} and the \ac{ieee39}. 
The systems' information were retrieved from Matpower \citep{zimmermann_matpower_2011}. 
We collected line admittance data, as well passive loads derived from demand data. 
For nodes equipped with units, we used the generation data as nominal active and reactive power setpoints $p^d$ and $q^d$. 
We considered unit models \eqref{eq:gfi_model} and \eqref{eq:sg_model} according to Table~\ref{tab:generators_and_nodes}. 
Thus, we obtain diverse setups: an \ac{sg} dominated \ac{ieee9}, a \ac{gfi} dominated \ac{ieee30} and the \ac{ieee39} with \acp{sg} only.
The global coordinate system is set to be equal to the local $dq$-coordinate system of the reference node according to Table~\ref{tab:generators_and_nodes}. 
The parameters, i.e., droop gains, time constants and inertia constants are varied arbitrarily around typical values.
\begin{table}[h]
    \vspace{0.7em}
	\centering
    \caption{Generators of the test systems.}
	\begin{tabular}{rccc}
		\toprule 
		System & $\mathcal{N}_{\text{gfi}}$ & $\mathcal{N}_{\text{sg}}$ & node for $\omega_{\text{ref}}, \varphi_{\text{ref}}$\\
        \midrule
        \ac{ieee9} & $\{1\}$ & $\{2,3\}$ & $1$\\
        \ac{ieee30} & $\{1,22,23,27\}$ & $\{2,13\}$ & $1$\\
        \ac{ieee39} & $\{\}$ & $\{30,31,\dots,39\}$ & $31$\\
		\bottomrule
	\end{tabular}
	\label{tab:generators_and_nodes}
\end{table} 

Recall from Section~\ref{sec:modeling} that the voltage amplitude $v_i$ at all nodes, as well as the frequency $\omega_i$ at \ac{sg} nodes and the power $p_i^m$ at \ac{gfi} nodes are measured. 
Measurements of voltage phase angles $\delta_i$ at the nodes were not available at all. 
The numerical integration method used with \acp{node} is \ac{rk4} with a fixed step-size equal to sampling step-size. 
We made use of the torchdiffeq package by \cite{torchdiffeq} which implements differentiable solvers in pytorch. 
For adjusting the weights of the models, we used the Adam optimizer \citep{kingma_adam_2017}. 

\subsection{Setup}
\subsubsection{Data generation}
\label{sec:data} 
The data was obtained from numerical simulations. 
For each system, we simulated three trajectories of length \SI{250}{\second} for training, validation and testing, as well as one trajectory of length  \SI{1010}{\second} for evaluation, at a sampling frequency of \SI{100}{Hz} each. 
As control inputs, we used random step changes in active power setpoints $p_i^d$, distributed equally within $\pm$\SI{0.2}{pu} around their nominal values. 
In the trajectories of length $\SI{250}{\second}$, they occur every \SI{5}{\second}, in the trajectories of length $\SI{1010}{s}$ every \SI{10}{\second}.
We simulated that each available output is affected by zero-mean Gaussian noise, such that each normalized signal has a signal to noise ratio of \SI{25}{dB}. 
\begin{remark}
Typically, a tertiary controller would change the setpoints on a timescale of several minutes. 
Such dynamic responses could realistically be captured over multiple setpoint changes, irrespective of their actual occurrence by selecting appropriate observation windows. 
\end{remark}

\subsubsection{Samples and datasets}
\label{sec:from_trajectory_to_training_samples} 
Based on the previously generated data, we formed four distinct datasets: For each system, we created one training, one validation, one testing and one evaluation dataset. 
The training dataset $\mathcal{D}^1$ was used to adjust the parameters $\theta$ of the model. 
The validation dataset $\mathcal{D}^2$ was used to monitor the training progress, i.e., to stop the training and to chose the best model over the epochs. 
For hyperparameter optimization, we relied on the error on the test dataset $\mathcal{D}^3$. 
Finally, the evaluation dataset $\mathcal{D}^4$ was used to assess the performance of the selected model.

\begin{figure}[h]
    \centering
    \input{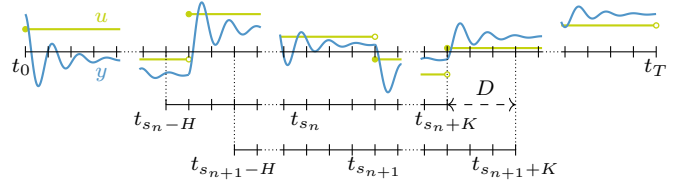}
    \caption{Cutting out two consecutive subsequences $n$ and $n+1$ with $D=3$ from a trajectory. Illustration inspired by~\cite{forgione_continuous_2021}.}
    \label{fig:datasets}
\end{figure}

To create the datasets $\mathcal{D}^m = \{(\xi_n, \eta_n)\}_{n=1}^{N_m}$, $m \in [1,4] \subset \mathbb{N}, N_m \in \mathbb{N}$, we constructed samples from simulated trajectories. 
We adopted the methodology presented in~\cite{forgione_continuous_2021} and cut out $N_m$ subsequences of length $H+K+1$ from $t_{s_n-H}$ to $t_{s_n+K}$  with a temporal distance $D$. 
Figure~\ref{fig:datasets} illustrates this procedure.
For $\mathcal{D}^1$, $\mathcal{D}^2$ and $\mathcal{D}^3$ we chose $H = 64$ historic time instants, a prediction horizon of $K = 64$ instants and $D=16$. 
For each, we cut out subsequences for $s_n = H + (n-1) \, D$ from the trajectories of length \SI{250}{\second} and formed samples $(\xi_n,\eta_n)$ as described in Section~\ref{sec:training_procedure}. 
We collected these samples into batches of size $256$ for training. 
To create $\mathcal{D}^4$, we used the trajectory of length  \SI{1010}{\second} and chose $H = 64$, $K = 500$ as well as  $D=1000$. 
Subsequences were cut out for $s_n = 1000 + (n-1) \, D$ such that for each resulting sample, $t_{s_n}$ aligns with the step changes in the control inputs at $t_{\text{step}_n} =  \SI{10}{\second}, \SI{20}{\second}, \dots,  \SI{1000}{\second}$. 
Thus, we used 100 samples in total for the evaluation.

\subsubsection{Hyperparameter Optimization}
\label{sec:hyperparameter_optimization} 

As hyperparameters, we considered the network sizes, i.e., the number of layers $L_f$ and hidden units $N_f$ of the MLP representing $f_{\theta_f}$, as well as the number of hidden channels $N_h$ used for the \ac{tcn} $h_{\theta_h}$. 
Note that we chose the minimum number of residual blocks $B$ such that $R \geq H$ and thus did not optimize the depth of the \ac{tcn}. 
Table~\ref{tab:ho_node} summarizes the search space. 
For Bayesian optimization of the hyperparameters, we used a sampler based on the tree-structured Parzen estimator \citep{bergstra_algorithms_2011} and a $25\%$-percentile pruner employing optuna \citep{akiba2019optuna}. 

\begin{table}[h]
    \vspace{0.7em}
	\centering
    \caption{Hyperparameter search space.}
	\begin{tabular}{rc}
		\toprule 
		Hyperparameter & Intervals/sets\\
        \midrule
        $f_{\theta_f}$ hidden layers & $L_f \in \{2^k \mid k = 0,1,2\}\subset \mathbb{N}_{\geq 0}$\\
		$f_{\theta_f}$ hidden neurons/layer & $N_f \in \{2^k \mid k = 5,6,\dots,10\}\subset \mathbb{N}_{\geq 0}$\\
        $h_{\theta_h}$ hidden channels & $N_h \in \{2^k \mid k = 5,6,\dots,10\} \subset \mathbb{N}_{\geq 0}$\\
        Learning rate & $\alpha \in [10^{-4},10^{-2}] \subset \mathbb{R}_{\geq 0}$\\
        $f_{\theta_f}$ activation function & $\sigma(\cdot) \in $\{Softplus, \acs{gelu}, \acs{silu}\}\\
		\bottomrule
	\end{tabular}
	\label{tab:ho_node}
\end{table}

\subsection{Results}
\label{sec:results}

In what follows, we analyze hyperparameter choice and prediction accuracy. 
We refer to our models as \ac{tcn}-\ac{anode}.
As a baseline, we use model \eqref{eq:partial_historic_anode} replacing \eqref{eq:partial_historic_augmentation} with simple partial input-layer augmentation that does not consider historic data, i.e., $x(t_s) = [{x^o}^T(t_s) ~ h_{\theta_h}^T(x^o(t_s))]^T$, where $h_{\theta_h}$ is an \ac{mlp}. 
A comparable approach has been used by \cite{xiao_feasibility_2023}.
We refer to these models as \ac{mlp}-\ac{anode}. 
We always use the best model found in the hyperparameter optimization for each power system and each augmentation method. 
We analyze the predictions of voltage magnitudes and frequencies on a previously unseen dataset. 
Note that $w_i$ was obtained using~\eqref{eq:frequency_droop_law} for all $i \in \mathcal{N}_{\text{gfi}}$. 

\subsubsection{Hyperparameter Choice} 
\label{sec:hyperparameter_choice}

Table~\ref{tab:ho_node_result} shows the the best hyperparameters found for each system for the \ac{tcn}-\ac{anode}. 
In general, we observe that for $f_{\theta_f}$ shallow networks with only one hidden layer perform best. 
Considering the number of hidden neurons per layer of $f_{\theta_f}$, the results vary between the systems. 
For the \ac{ieee9} less hidden neurons than for the \ac{ieee30} and the \ac{ieee39} were required to capture the dynamics accurately. 
The same holds for the number of hidden channels of the \ac{tcn} $h_{\theta_h}$. 
However, we would like to emphasize, that for all systems, the error barely changed when double or half the number of hidden neurons or channels were chosen. 
The learning rate was found to be in typical ranges for \ac{nn} training, but it was necessary to adjust it accordingly for the respective network size. 
From all the smooth activation functions, we found that the \ac{silu} performed best across all systems. 

\begin{table}[h]
    \vspace{0.7em}
	\centering
    \caption{Best Hyperparameters for all systems with the \ac{tcn}-\ac{anode}.}
	\begin{tabular}{rccc}
		\toprule
	    System & \ac{ieee9} & \ac{ieee30} & \ac{ieee39} \\
        \midrule
        $f_{\theta_f}$ hidden layers & $1$ & $1$ & $1$ \\
		$f_{\theta_f}$ hidden neurons & $128$ & $1024$ & $512$\\
        $h_{\theta_h}$ hidden channels & $32$ & $64$ & $64$ \\
        Learning rate & $5.22 \cdot 10^{-3}$& $4.43 \cdot 10^{-3}$ & $1.01\cdot 10^{-3}$\\
        $f_{\theta_f}$ activation fct. & \acs{silu} & \acs{silu} & \acs{silu} \\
		\bottomrule
	\end{tabular}
	\label{tab:ho_node_result}
\end{table}

\subsubsection{Prediction Accuracy}
\label{sec:longterm_prediction_accuracy} 

\begin{figure}
    \centering

\tikzset{external/export next=true}

\pgfplotsset{resultsPlot/.style={%
    clip = false,
    minor x tick num=1,
    grid=both,
    grid style={draw=black!25},
    major tick length=0pt,
    minor tick length=0pt,
    axis lines = left,
    axis line style= {-, draw opacity=0.0},
    y tick label style={
        /pgf/number format/.cd,
            scaled y ticks = false,
            fixed,
            precision=3,
        /tikz/.cd
        },
    height = 4cm,
		xtick = {20, 21, 22, 23, 24, 25},
		xmin = 20, 
		xmax = 25,
		clip=true, 
    width=8.5cm,
    legend columns=3,
    legend style={
      at={(0.5, 1.05)},
      anchor=south,
      draw=none,
      fill=none,
      legend cell align=left,
      /tikz/every even column/.append style={column sep=2.5mm},
      inner sep=0pt,
      legend cell align={left},
      },
  	} 
}

\pgfplotsset{resultsPlotLong/.style={%
    clip = false,
    minor x tick num=1,
    grid=both,
    grid style={draw=black!25},
    major tick length=0pt,
    minor tick length=0pt,
    axis lines = left,
    axis line style= {-, draw opacity=0.0},
    y tick label style={
        /pgf/number format/.cd,
            scaled y ticks = false,
            fixed,
            precision=3,
        /tikz/.cd
        },
    height = 4cm,
		clip=false,
    width=8.5cm,
    legend columns=3,
    legend style={
      at={(0.5, 1.05)},
      anchor=south,
      draw=none,
      fill=none,
      legend cell align=left,
      /tikz/every even column/.append style={column sep=2.5mm},
      inner sep=0pt,
      legend cell align={left},
      },
  	}
}
\def\plottestcase{ieee30}
\def\plotmodel{restcn}

\def\plotcolormeas{black!50}
\def\plotcolortrue{black}

\def\linewidthmeas{0.7}
\def\linewidthtrue{0.7}
\def\linewidthpred{1.25}

\ifnum\colortheme=1
  \def\plotcolorpredone{Hellblau}
  \def\plotcolorpredtwo{Gelbgrue}
  \def\plotcolorpredthree{Hellgrue}
\else\ifnum\colortheme=2  
  \def\plotcolorpredone{cbBlue}
  \def\plotcolorpredtwo{cbGreen}
  \def\plotcolorpredthree{cbOrange}
\fi\fi

\begingroup  
  \def\temp{ieee9}%
  \ifx\plottestcase\temp
    \gdef\plotnodeone{1}%
    \gdef\plotnodetwo{2}%
    \gdef\plotnodethree{3}%
  \fi 

  \def\temp{ieee30}%
  \ifx\plottestcase\temp
    \gdef\plotnodeone{1}%
    \gdef\plotnodetwo{13}%
    \gdef\plotnodethree{27}%
  \fi
 
  \def\temp{ieee39}%
  \ifx\plottestcase\temp
    \gdef\plotnodeone{30}%
    \gdef\plotnodetwo{31}%
    \gdef\plotnodethree{32}%
  \fi
\endgroup
 
\begin{tikzpicture}[font=\footnotesize]

\draw[draw=none, fill=none] (-1.45, 2.55) rectangle (7.2, -3.8); 

\begin{axis}[%
    resultsPlot,
    xlabel = {},
    xticklabels = {~},
    ylabel={Voltage [pu]},
    minor y tick num=1,
]
  \foreach \x in {\plotnodeone,\plotnodetwo,\plotnodethree}{
    \def\fileone{\plottestcase \plotmodel measstep1.csv}
    \addplot[line width=\linewidthmeas pt, color=\plotcolormeas, opacity=0.35, forget plot] table [x=time, y=v\x, col sep=comma, each nth    point=1, ]{data/\fileone};
  }
  \foreach \x in {\plotnodeone,\plotnodetwo,\plotnodethree}{
    \def\filetwo{\plottestcase \plotmodel truestep1.csv}
    \addplot[line width=\linewidthtrue pt, color=\plotcolortrue, forget plot, opacity=0.75] table [x=time, y=v\x, col sep=comma, each nth    point=1, ]{data/\filetwo};  
  }
  \foreach \x in {\plotnodeone,\plotnodetwo,\plotnodethree}{
    \def\filethree{\plottestcase \plotmodel predstep1.csv}
    \ifnum\x=\plotnodeone
      \addplot[line width=\linewidthpred pt, color=\plotcolorpredone, densely dashdotted] table [x=time, y=v\x, col sep=comma, each nth    point=1, ]{data/\filethree};
    \else\ifnum\x=\plotnodetwo
      \addplot[line width=\linewidthpred pt, color=\plotcolorpredtwo, densely dashdotted] table [x=time, y=v\x, col sep=comma, each nth    point=1, ]{data/\filethree};
    \else\ifnum\x=\plotnodethree
      \addplot[line width=\linewidthpred pt, color=\plotcolorpredthree, densely dashdotted] table [x=time, y=v\x, col sep=comma, each nth    point=1, ]{data/\filethree};
    \fi\fi\fi
  } 
  \addlegendentry{$\hat{v}_{\plotnodeone}$, $\hat{\omega}_{\plotnodeone}$}
  \addlegendentry{$\hat{v}_{\plotnodetwo}$, $\hat{\omega}_{\plotnodetwo}$}
  \addlegendentry{$\hat{v}_{\plotnodethree}$, $\hat{\omega}_{\plotnodethree}$}
  \addlegendimage{line width=\linewidthmeas  pt, color=\plotcolormeas, opacity=0.5}
  \addlegendentry{Measured signal}
  \addlegendimage{line width=\linewidthtrue pt, color=black}
  
  \addlegendentry{True signal}

\end{axis}

\begin{axis}[%
    yshift = -2.75cm,
    resultsPlot,
    xlabel = {Time [s]},  
    ylabel={Ang. vel. [rad/s]},
    minor y tick num=1,
]
  \foreach \x in {\plotnodeone,\plotnodetwo,\plotnodethree}{
    \def\filefour{\plottestcase \plotmodel measstep1.csv}
    \addplot[line width=\linewidthmeas pt, color=\plotcolormeas, opacity=0.35, forget plot] table [x=time, y=omega\x, col sep=comma, each nth    point=1, ]{data/\filefour};
  }
    \foreach \x in {\plotnodeone,\plotnodetwo,\plotnodethree}{
    \def\filefive{\plottestcase \plotmodel truestep1.csv}
    \addplot[line width=\linewidthtrue pt, color=\plotcolortrue, forget plot, opacity=0.75] table [x=time, y=omega\x, col sep=comma, each nth    point=1, ]{data/\filefive}; 
  }
  \foreach \x in {\plotnodeone,\plotnodetwo,\plotnodethree}{
    \def\filesix{\plottestcase \plotmodel predstep1.csv}
    \ifnum\x=\plotnodeone
      \addplot[line width=\linewidthpred pt, color=\plotcolorpredone, densely dashdotted] table [x=time, y=omega\x, col sep=comma, each nth    point=1, ]{data/\filesix};
    \else\ifnum\x=\plotnodetwo
      \addplot[line width=\linewidthpred pt, color=\plotcolorpredtwo, densely dashdotted] table [x=time, y=omega\x, col sep=comma, each nth    point=1, ]{data/\filesix};
    \else\ifnum\x=\plotnodethree
      \addplot[line width=\linewidthpred pt, color=\plotcolorpredthree, densely dashdotted] table [x=time, y=omega\x, col sep=comma, each nth    point=1, ]{data/\filesix};
    \fi\fi\fi
  } 
\end{axis}

\end{tikzpicture}
    \caption{Voltage amplitude and frequency responses for the \ac{ieee30} at nodes 1,13 and 27 to a step change in $p_i^d$ for all units at $t_s = \SI{20}{s}$. }
    \label{fig:simulation}
\end{figure}
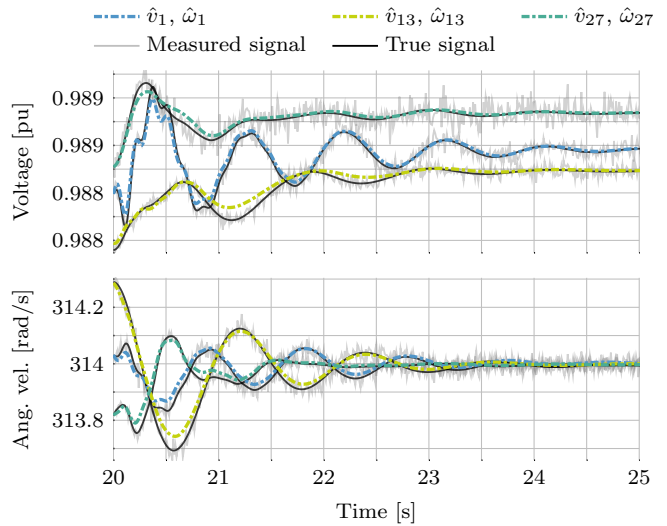
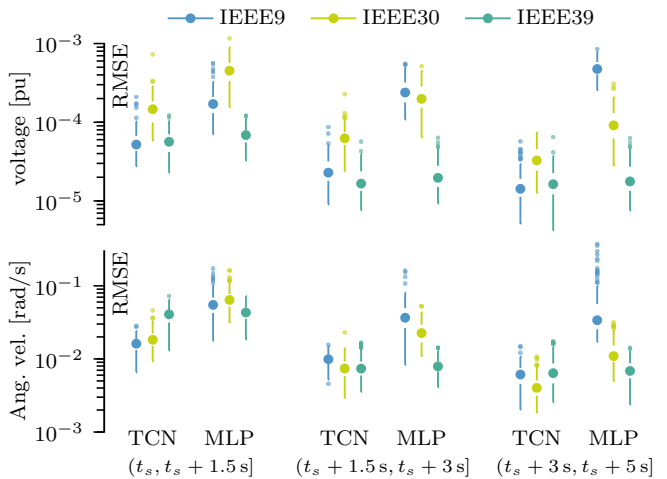
\begin{figure}[h]
    \centering

\tikzset{external/export next=true}

\pgfplotsset{linestyle boxplot/.style={%
  boxplot = {%
    every box/.style={draw=none, fill=none},
    whisker extend=0,
    draw direction=y,
    },
    mark=*,
    every mark/.append style={mark size=0.7pt, line width=0pt, opacity=0.6, fill=#1}, draw=#1,
    boxplot/draw/median/.code={%
          \draw[mark size=1.5pt, /pgfplots/boxplot/every median/.try]
          \pgfextra
          \pgftransformshift{
            \pgfplotsboxplotpointabbox
              {\pgfplotsboxplotvalue{median}}
              {0.5}
          }
          \pgfsetfillcolor{#1}
          \pgfuseplotmark{*}
          \endpgfextra
        ;
      },
  },
}

\ifnum\colortheme=1
  \def\plotcolorpredone{Hellblau}
  \def\plotcolorpredtwo{Gelbgrue}
  \def\plotcolorpredthree{Hellgrue}
\else\ifnum\colortheme=2
  \def\plotcolorpredone{cbBlue}
  \def\plotcolorpredtwo{cbGreen}
  \def\plotcolorpredthree{cbOrange}
\fi\fi

\begin{tikzpicture}[font=\footnotesize]

  \begin{semilogyaxis}[
    myPlot,
    clip=false,
    height = 40mm,
    width = 85mm,
    ymin = 5e-6,
    ymax = 1e-3,
    line width=0.7pt,
    x axis line style={white},
    xtick style={draw=none},
    xtick = {},
    xticklabels={},
    xticklabel style={draw=none},
    clip=false,
    ylabel = {RMSE }, 
    y label style={at={(0, 1)}, anchor=east, inner sep=0pt},
    legend columns=3,
    legend style={
      at={(0.5, 1.05)},
      anchor=south,
      draw=none,
      fill=none,
      legend cell align=left,
      /tikz/every even column/.append style={column sep=2.5mm},
      inner sep=0pt,
      legend cell align={left},
      },
    ]
  
  \addlegendimage{line legend, line width = 0.7pt, mark=*, mark size=1.5pt, color=\plotcolorpredone};
  \addlegendentry{IEEE9};

  \addlegendimage{line legend, line width = 0.7pt, mark=*, mark size=1.5pt, color=\plotcolorpredtwo};
  \addlegendentry{IEEE30};
  
  \addlegendimage{line legend, line width = 0.7pt, mark=*, mark size=1.5pt, color=\plotcolorpredthree};
  \addlegendentry{IEEE39};

  \foreach \i in {0,1,2} {
    \addplot [linestyle boxplot=\plotcolorpredone, boxplot/draw position={1+5*\i - 0.15mm}] table[y index= \i, col sep=comma]{data/ieee9restcnrmsebracketsv.csv};
  }

  \foreach \i in {0,1,2} {
    \addplot [linestyle boxplot=\plotcolorpredtwo, boxplot/draw position={1+5*\i}] table[y index= \i, col sep=comma]{data/ieee30restcnrmsebracketsv.csv};
  }
  \foreach \i in {0,1,2} {
    \addplot [linestyle boxplot=\plotcolorpredthree, boxplot/draw position={1+5*\i + 0.15mm}] table[y index= \i, col sep=comma]{data/ieee39restcnrmsebracketsv.csv};
  }

  \foreach \i in {0,1,2} {
    \addplot [linestyle boxplot=\plotcolorpredone, boxplot/draw position={3+5*\i - 0.15mm}] table[y index= \i, col sep=comma]{data/ieee9mlprmsebracketsv.csv};
  }

  \foreach \i in {0,1,2} {
    \addplot [linestyle boxplot=\plotcolorpredtwo, boxplot/draw position={3+5*\i}] table[y index= \i, col sep=comma]{data/ieee30mlprmsebracketsv.csv};
  }
  \foreach \i in {0,1,2} {
    \addplot [linestyle boxplot=\plotcolorpredthree, boxplot/draw position={3+5*\i + 0.15mm}] table[y index= \i, col sep=comma]{data/ieee39mlprmsebracketsv.csv};
  }

  \node[anchor=north east, rotate=90] at (rel axis cs: -0.23, 0.9) {voltage [pu]};

  \end{semilogyaxis}

  \begin{semilogyaxis}[
  yshift=-27.5mm,
  myPlot,
  clip=false,
  height = 40mm,
  width = 85mm,
  ymin = 1e-3,
  ymax = 3e-1,
  line width=0.7pt,
  x axis line style={white},
  xtick style={draw=none},
  xticklabels={TCN, , MLP, , , TCN, , MLP, , , TCN, , MLP},
  xtick = {1,2,3,4,5,6,7,8,9,10,11,12,13},
  xticklabel style={yshift=4mm}, 
  extra x ticks      = {2,7,12},
  extra x tick labels={\scriptsize{$(t_{s}, t_{s} + \SI{1.5}{\second}]$}, \scriptsize{$(t_{s}+ \SI{1.5}{\second}, t_{s} + \SI{3}{\second}]$}, \scriptsize{$(t_{s}+ \SI{3}{\second}, t_{s} + \SI{5}{\second}]$}},
  extra x tick style={
    tick label style={
      yshift=-5mm,  
      anchor=north,
    },
    major tick length=0pt, 
  },
  clip=false,
  ylabel = {RMSE },
  y label style={at={(0, 1)}, anchor=east, inner sep=0pt},
  ]

  \foreach \i in {0,1,2} {
    \addplot [linestyle boxplot=\plotcolorpredone, boxplot/draw position={1+5*\i - 0.15mm}] table[y index= \i, col sep=comma]{data/ieee9restcnrmsebracketsw.csv};
  }

  \foreach \i in {0,1,2} {
    \addplot [linestyle boxplot=\plotcolorpredtwo, boxplot/draw position={1+5*\i}] table[y index= \i, col sep=comma]{data/ieee30restcnrmsebracketsw.csv};
  }
  \foreach \i in {0,1,2} {
    \addplot [linestyle boxplot=\plotcolorpredthree, boxplot/draw position={1+5*\i + 0.15mm}] table[y index= \i, col sep=comma]{data/ieee39restcnrmsebracketsw.csv};
  }

  \foreach \i in {0,1,2} {
    \addplot [linestyle boxplot=\plotcolorpredone, boxplot/draw position={3+5*\i - 0.15mm}] table[y index= \i, col sep=comma]{data/ieee9mlprmsebracketsw.csv};
  }

  \foreach \i in {0,1,2} {
    \addplot [linestyle boxplot=\plotcolorpredtwo, boxplot/draw position={3+5*\i}] table[y index= \i, col sep=comma]{data/ieee30mlprmsebracketsw.csv};
  }
  \foreach \i in {0,1,2} {
    \addplot [linestyle boxplot=\plotcolorpredthree, boxplot/draw position={3+5*\i + 0.15mm}] table[y index= \i, col sep=comma]{data/ieee39mlprmsebracketsw.csv};
  }

  \node[anchor=north east, rotate=90] at (rel axis cs: -0.23, 1.0) {Ang. vel. [rad/s]};

  \end{semilogyaxis}

\end{tikzpicture}
    \caption{Box plots of the prediction \acs{rmse} of 100 step responses for all systems over a horizon of \SI{5}{s} splitted into different time intervals.
    }
    \label{fig:boxplot}
\end{figure}

To analyze the prediction accuracy of the learned models we use the evaluation dataset $\mathcal{D}^4$. 
For all units, each sample is associated with a prediction of $\omega_i$ and $v_i$ to a step change in $p_i^d$  at time $t_{s_n}$. 
Figure~\ref{fig:simulation} shows an example trajectory of such a response for the \ac{ieee30} at nodes 1, 13 and 27 for sample $n=2$. 
A visual inspection indicates that frequency and voltage dynamics are captured well by the learned model. 
Transient errors at the beginning of the response are small, emphasizing the quality of the learned initial state. 
Figure~\ref{fig:boxplot} visualizes the model accuracy in a quantitative manner using the \ac{rmse} associated with all 100 samples. 
The \ac{rmse} is calculated for three time intervals $(t_{s}, t_{s} + \SI{1.5}{\second}]$, $(t_{s} +  \SI{1.5}{\second}, t_{s} +  \SI{3}{\second}]$ and $(t_{s} +  \SI{3}{\second}, t_{s} +  \SI{5}{\second}]$ capturing short-term, medium-term and long-term accuracy, respectively, ranging from highly dynamic to quasi steady-state conditions. 

Our novel \ac{tcn}-\ac{anode} reaches strong steady state accuracy. 
The slightly decreased short-term  accuracy most likely originates from small errors in the augmented part of $x(t_{s})$ as well as noise in $x^o(t_{s})$ which directly translates into errors in initial state $x(t_{s})$. 
However, these errors fade over time as the system converges to steady state.  
Meanwhile, the \ac{mlp}-\ac{anode} shows significantly larger errors in all time brackets and fails to capture the dynamics accurately for the \ac{ieee9} and \ac{ieee30} test cases. 
A visual inspection of the \ac{mlp}-\ac{anode} trajectories of the \ac{ieee9} test case showed that the resulting trajectories drifted, eventually leading to instability over a longer time horizon. 
This shows that the \ac{tcn}-\ac{anode} allows learn appropriate latent representation of phase angles resulting in an overall accurate identification, while the \ac{mlp}-\ac{anode} fails to achieve this goal. 


\section{Conclusion}
\label{sec:conclusion}

In this paper, we proposed a novel structure based on \acp{anode} to identify power system dynamics. 
We employed a \ac{tcn} to learn latent representations for phase angles based on historic observations. 
The approach allows for accurate identification without the necessity of phase angle measurements and outperforms simpler augmentation techniques, that are widely employed. 
In the future, we intend to combine known and unknown parts for the system identification, as well as architectural novelties like graph \acp{node} to identify power system dynamics.

\bibliography{literature}

\end{document}